\documentclass[12pt]{article}
\usepackage[pdftex]{graphicx}
\usepackage{color}
\usepackage{amsmath}
\usepackage{lineno}              

\begin{document}

\newsavebox{\savepar}
\newenvironment{boxit}{\begin{lrbox}{\savepar}
\begin{minipage}[b]{7in}}
{\end{minipage}\end{lrbox}\fbox{\usebox{\savepar}}}

\newcommand{\be}{\begin{equation}}
\newcommand{\ee}{\end{equation}}
\newcommand{\red}{\color{red}}
\newcommand{\blue}{\color{blue}}
\newcommand{\black}{\color{black}}
\parskip 10pt
\parindent 20pt

\title{ {\bf Life After Earth}}

\vspace{-1.0cm}

\author{Caren Marzban$^{1,2}$\thanks{Corresponding Author:
marzban@stat.washington.edu}, Raju Viswanathan$^3$, Ulvi Yurtsever$^{4}$ \\
$^1$ Applied Physics Laboratory, $^2$ Department of Statistics \\
Univ. of Washington, Seattle, WA, 98195 USA \\
$^3$ Technome, LLC, 8107 Colmar Drive, Clayton MO 63105 \\
$^4$ MathSense Analytics, 1273 Sunny Oaks Circle, Altadena, CA 91001 USA\\
}
\date{ }
\maketitle
\vspace{-1.0cm}

\begin{abstract}
A recent study reported that there is evidence life may have
originated prior to the formation of the Earth. That conclusion
was based on a regression analysis of a certain data set involving
evolution of functional genome size across major phyla. Here it is 
shown that if measurement errors and ``confidence" intervals are taken
into account, then the regression analysis of the same data set 
leads to conclusions that allow for life to have appeared after the 
formation of the Earth.  
\end{abstract}

\section{Introduction}

Recently, Sharon and Gordon (2013) - hereafter SG - reported an
analysis of data on the evolution of genetic complexity during
the history of life on Earth. As a measure of genetic complexity SG
use the functional genome size of major phylogenetic lineages, whose
logarithms become the $y$-axis values of their data set, with the estimated
dates of the transitions where these lineages first originated being
the $x$-axis values.  
They performed regression on the data (on $y$ vs. $x$), and proposed
that the $x$-intercept of the fit provides an estimate for the age of life.

The work was criticized on many levels, ranging from the manner in which the
data 
was produced, to the way in which the data was analyzed. A fundamental problem 
is the paucity of data over the first 2 billion years or so of Earth's history, 
resulting in large uncertainties in functional genome size at specific times. 
For instance, for prokaryotes, the size of the functional genome is guessed
from 
the smallest present-day prokaryote genome. Exactly when this genome size
evolved 
is a matter of conjecture; although an approximate date can be estimated from 
molecular clock type evolution rates based on more recent organisms, as some 
reviewers have pointed out (see, for instance, Sharov 2006), rates of increase
of functional genome size could have been very different in the distant past.  
Fitting the data with an extrapolation based on a single, fixed rate of increase
could lead to possibly incorrect conclusions. 
Likewise, the use of only coding regions of the genome as a measure of genome 
complexity has been pointed out as a potential problem, as non-coding regions 
could play a regulatory role and the associated complexity is unaccounted for 
when only coding regions are measured. Thus estimating genome complexity 
of extinct organisms based on an uncertain estimate of functional genome size 
of present-day organisms could be doubly flawed.  

In addition to all of the above criticisms, there are additional concerns
over the statistical analysis in SG.  First, and foremost, is the way in 
which the regression fit is used to extrapolate far beyond the range of $x$ 
values appearing in the data. It is well known that extrapolation can lead to
misleading conclusions (Perrin 1904), and so, any conclusions regarding the age of
life, based on extrapolation, should be considered with extreme caution. A second 
aspect of the SG regression fit is that it does not incorporate ``confidence" 
intervals. Here, the term ``confidence" is used loosely; the interval actually 
computed in this study is the {\it prediction interval} (see below). 
The inclusion of such intervals can lessen the misleading impacts of extrapolation,
because prediction intervals generally widen as one moves away from the
mean of the data. Then, the $x$-intercept is accompanied by
a relatively wide range of values, all of which are equally likely
values for the age of the life. In other words, inclusion of prediction
intervals can further mitigate misleading conclusions. Another limitation of 
the SG regression analysis is that it does not account for uncertainty in 
the dates at which the transitions occurred (i.e., the $x$-values of the data), 
also known as {\it measurement errors.} As explained here, measurement errors 
generally reduce the slope of the regression fit, and consequently increase the 
value of the $x$-intercept. As such, measurement errors lead to overestimates 
for the age of life.

In this paper, a simple measurement error model is developed, and rudimentary
prediction intervals are produced. First, an attempt is made to 
estimate the 
measurement errors, and then, it is shown that a measurement error model of
the data leads to conclusions that are consistent with life having formed
around 4.5-billion years ago. In short, we find that when the regression 
analysis includes prediction intervals and incorporates measurement errors, 
then the data used by SG provide no evidence to support the claim that 
life must have formed prior to the formation of the Earth.

\section{Regression Effect}

Consider a scatterplot of $y$ vs. $x$, displaying some amount of
association between the two variables (e.g., Figure 1). It is well 
known that as the spread of the data increases, the slope of a
least-squares fit approaches zero. This effect is known by
a variety of names, including {\it the regression effect} (Bland and Altman 1994).
It is demonstrated in Figure 1, where the black circles have less scatter
than the red circles. The straight lines are the ordinary least-squares
fits to respective data. It can be seen that increasing scatter leads to
lower values of the slope. (In this particular case, the red circles have
been generated by adding error to the $x$-values of the black circles.)

\centerline{\includegraphics[height=3in,width=4in]{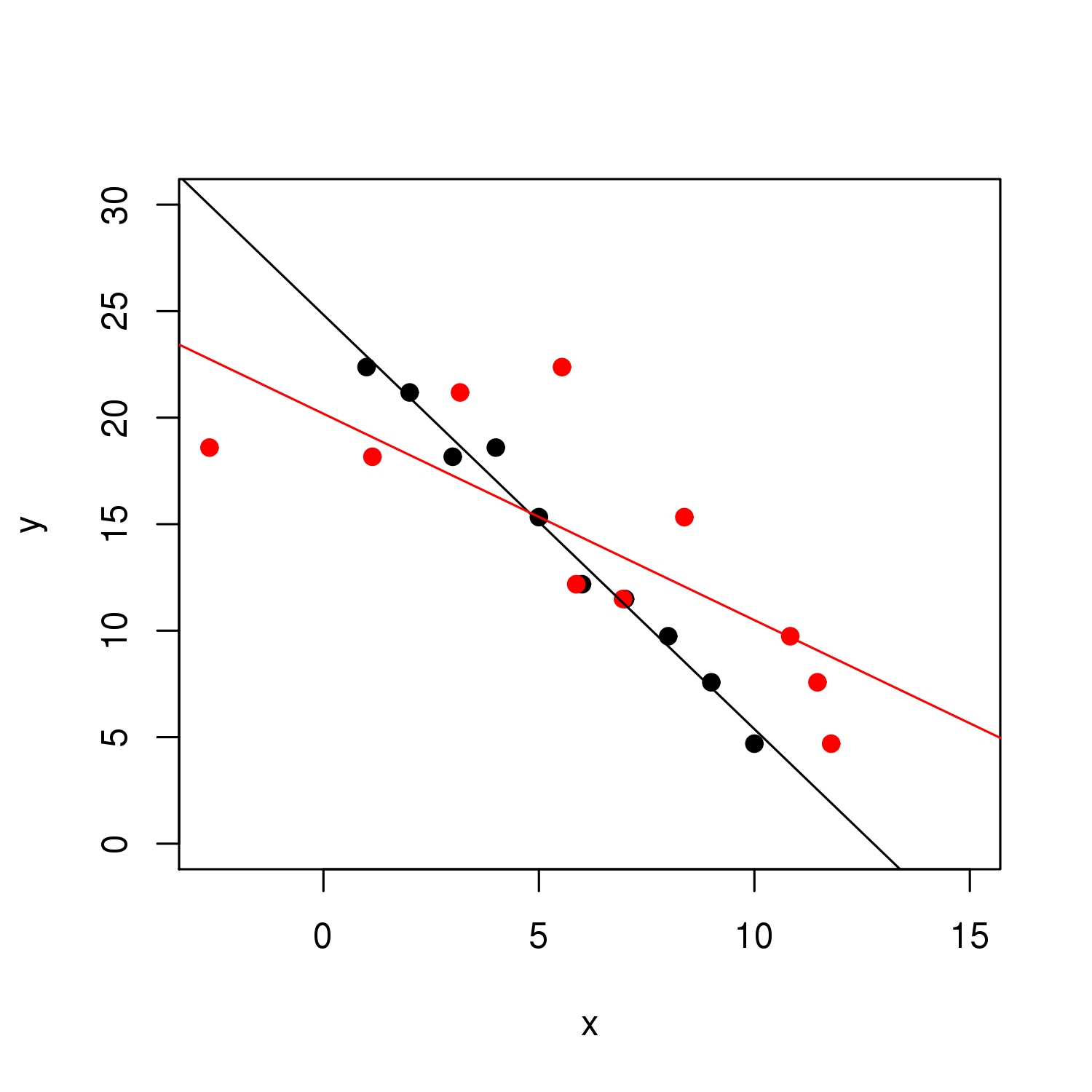} }

The mathematics underlying the regression effect is straightforward.
It is easy to show 
\be
\frac{\hat{y}(x) - \overline{y}}{s_y} = r (\frac{x-\overline{x}}{s_x}),
\ee
where $\hat{y}(x)$ is the predicted/fitted value, $\overline{x}$
and $\overline{y}$ are the sample mean of $x$ and $y$, respectively,
and $s_x$, $s_y$ are their sample standard deviations. The quantity
$r$ is Pearson's correlation coefficient, and it measures the
amount of scatter on the scatterplot. As $r$ approaches zero
(from either side), the predicted value $\hat{y}(x)$ tends
to the sample mean of $y$. Indeed, this ``regression to the
mean" is the reason why the least-squares fit is called 
regression (Galton 1886). In summary, as the amount of scatter in
the scatterplot of $y$ vs. $x$ increases, the least-square fit
converges to a horizontal line with slope zero, and y-intercept
equal to $\overline{y}$. 

\section{Regression Dilution}

The aforementioned scatter may be due to errors in $x$, in $y$, or both.
In the most common form of regression, the predictor $x$ is assumed to 
be error-free, and only the response is assumed to be subject to errors. 
Measurement error models (Buonaccorsi 2010; Fuller 1987) are designed to allow
for both $x$ and $y$ to be subject to errors. Consequently, as expected from
the previous example, measurement errors tend to ``flatten" the least-squares 
line - a phenomenon called ``regression dilution" (Buonaccorsi 2010; Fuller 1987). 
Moreover, if the measurement errors can be estimated, then one can undo the 
dilution. 

A simple measurement error model is as follows: 
Let $(X_i,Y_i), i = 1, ..., a$, denote the true, error-free, values of two 
continuous random variables, satisfying the relation
\be Y_i = \alpha^* + \beta^* \; X_i \;.  \ee
The corresponding observed values $(x_i,y_i)$ can then be written as
\be x_i = X_i + \omega_i \;\;\; , \;\;\;\; y_i = Y_i + \epsilon_i \;, \ee
where $\omega_i$ and $\epsilon_i$ are the measurement error in $X$ and the 
error in $Y$, respectively. For simplicity, it is assumed that both are normally 
distributed with zero mean, and variances given by $\sigma_w^2$ and 
$\sigma^2_\epsilon$.  I.e., $\omega_i \sim N(0,\sigma^2_w)$ and 
$\epsilon_i \sim N(0,\sigma^2_{\epsilon})$. In the {\it functional model}
the $X$ is assumed fixed (non-random), but in the {\it structural model}
$X$ is assumed to be a random variable (Buonaccorsi 2010; Fuller 1987). In the 
former, the $a$
values of $X_i$ are assumed to be fixed quantities, while in the
latter they are considered to be a random sample taken from a population
(or a distribution). 
The latter is adopted here, because it is more appropriate for the
problem at hand. Again, for simplicity, one assumes 
$X_i \sim N(\mu, \sigma^2_b)$. \footnote{The
subscripts ``b" and ``w" are motivated by ``between-group" and
``within-group" variances - language common to the analysis of
variance formulation of regression (Montgomery 2009).}

If one mistakenly ignores measurement errors (in $X$), and instead performs 
regression on $(x_i,y_i)$, i.e.,
\be 
y_i = \alpha + \beta \, x_i + \epsilon_i \;,
\ee
then it can be shown that the least-squares estimate of the 
regression slope and y-intercept are given by (Fuller 1987; Draper and Smith 1998)
\be \beta = \beta^* / \lambda\; \;, \;\;  
\alpha = \overline{y} - (\beta^* / \lambda)\, \overline{x} \;\;,\ee
with
\be \lambda = 1 + \frac{\sigma_w^2}{\sigma_b^2} \;. \ee
Given that $\lambda>1$, it follows that $\beta < \beta^*$, i.e., the
slope is ``diluted" relative to the slope that would have been obtained
if measurement errors were zero.  Said differently, measurement errors tend to 
``flatten" the least-squares fit, and therefore, lead to an overestimate 
of the $x$-intercept. In the framework of SG, then,
measurement errors lead to an overestimate for the age of life.  
In a measurement error model of SG's data, the corrected
$x$-intercept $\overline{x} - \overline{y}/(\beta\, \lambda)$ estimates
the age of life.  

Equation (5) implies that one can correct this effect, by simply multiplying the
observed regression coefficient $\beta$ by $\lambda$. 
In other words, the quantity ($\beta\,\lambda$) is an estimator of $\beta^*$. 
Similarly, the least square estimate of $\alpha$ is 
$(\overline{y} - (\beta \, \lambda) \, \overline{x})$. In order to make
these corrections, however, one must estimate $\lambda$.

Frost and Thompson (2000) discuss six methods for estimating $\lambda$,
and the corresponding variance. One of the simpler methods examined there 
identifies $\lambda$ as the inverse of the intraclass correlation coefficient 
(also known as the reliability ratio). One advantage of that estimator
is that its variance has a simple expression:
\be 
\frac{(\lambda^2-1)^2}{a} \; .
\ee
Although in the next section an attempt is made to estimate $\lambda$ itself,
the main focus of the study is to consider the ``inverse problem" of finding a 
range of $\lambda$ values which lead to $x$-intercepts consistent with 4.5 
billion years as the age of life. 

It is not necessary to find a specific $\lambda$ value which
leads to an $x$-intercept of 4.5 billion years. A regression fit
whose prediction interval includes an $x$-intercept of 4.5 billion
years is sufficient, in the sense that it does not contradict
the null hypothesis that life began after the formation of the Earth.
To that end, we supplement all of the above estimates with
prediction intervals. In order to construct such an interval, one must 
compute the variance for the corrected regression slope, a quantity which
has been derived by Frost and Thompson (2000):
\be 
V[(\beta\,\lambda)] = \lambda^2 V[\beta] + 
\frac{1}{a} (\beta^2 + V[\beta]) (\lambda^2-1)^2 \,.
\ee
where Eq. (7) has been used.

There is an ambiguity in whether the appropriate interval for this 
problem is a confidence interval or a prediction interval (Ryan 1997). The former 
is designed to cover the true conditional mean 
of $y$, given $x$, a certain percentage of time, e.g., 95\%. The latter 
is designed to cover a single prediction of $y$, a certain percentage of time.
By construction, the prediction interval is wider than the confidence interval. 
Here, a prediction interval is considered, because our interest is in the
$x$-intercept, which corresponds to a single prediction of $y$.
The choice between the two intervals is of secondary importance. What is more 
important than the choice of the two intervals
is that {\it some} interval must be considered. 

The construction of prediction intervals in measurement error models is
itself a complex issue and is considered by Buonaccorsi (1995). One relatively
simple 95\% prediction interval is given by 
$\hat{y}(x) \pm 1.96 \sigma_{pe}$, where $\sigma^2_{pe}$ is the variance
of the prediction error, given by
\be 
  \sigma^2_{pe} = \sigma^2_{\epsilon} + \frac{\sigma_{\epsilon}^2}{a} + 
                   (X-\overline{X})^2 V[\beta \lambda] +
            [(\beta \, \lambda)^2 + V[\beta \, \lambda]]\, \sigma_b^2 \;\; ,
\ee
where $V[(\beta \, \lambda)]$ is given by Eq. 8, and $\sigma_{\epsilon}^2$ is
estimated by the variance of the residuals. This is the expression 
derived in Buonaccorsi (1995) for the special case where the value of $X$ at 
which the prediction is made is a known (non-random) quantity. The first three
terms on the right-hand side of Eq. (9) are the variance of the prediction
error in the error-free case (Draper and Smith 1998); the last term
is the result of measurement errors.

\section{Estimating Measurement Errors}

One may wonder what is a typical value of $\lambda$ for the data at hand.
For that, $\sigma_b$ and $\sigma_w$ must be estimated. To that end,
consider a situation where each $X_i$ is measured $n$ times.
Denoting the resulting data as $x_{ij}, i = 1, ..., a, \, j = 1, ..., n$,
it is known that unbiased estimates of $\sigma_b^2$ and $\sigma_w^2$ are
\be
(\frac{s_b^2}{n} - \frac{s_w^2}{n}) \;\;, \;\;  s_w^2 \;,
\ee
respectively, with $s_b^2$ and $s_w^2$ defined as
\be
s_b^2 = \frac{n}{a-1} \sum_i^a (\overline{x_{i.}} - \overline{x_{..}})^2 \;\;, \;\;
s_w^2 = \frac{1}{a(n-1)} \sum_{i,j}^{a,n} (x_{ij} - \overline{x_{i.}})^2 \;\;.
\ee
where an overline denotes averaging over the index with a dot (Montgomery 2009).
For large $n$, the quantity $s_b^2/n$ converges to the sample variance
of the $X_i$, i.e., $s_X^2 = \frac{1}{a-1} \sum_i^a (X_i - \overline{X})^2$,
which in turn can be estimated with the sample variance of the $x_i$. For the
data at hand, then, $s_b^2/n \sim 1.86$ billion years. In the large-$n$
limit, the term $s_w^2/n$ converges to zero, because in that limit
$s_w^2$ itself converges to the constant $\sigma_w^2$. Therefore, asymptotically,
$\sigma_b \sim \sqrt{1.86} \sim 1.36$.

The within-group standard deviation $s_w$ reflects the spread in values or
uncertainty of the dates of appearance of the respective functional genomes
(e.g., prokaryote, eukaryote, worms, fish, mammals) used in the data analysis.
While the statistical analysis presented here assumes that the $a$
measurements all have common variance (i.e., homoscedastic), in reality the
uncertainty in the time
of appearance of a functional genome increases from present to past. Thus the
largest errors or uncertainties are found in the oldest functional genome
considered. As an example of dating uncertainty, while the earliest mammals
are believed to have arisen about 225 million years ago based on early fossils
(Rose 2006), molecular clock studies based on genomes place mammalian origins
around 100 million years ago (Dawkins 2005). There is thus an uncertainty of
the order of 100 million years or more in setting the time of the mammalian
functional genome. The origin of eukaryotes has been identified to lie in the
time interval between 2.3 billion and 1.8 billion years ago, thus with an
uncertainty of 250 million years around the mean estimate of 2.05 billion
years ago (Seilacher, Bose, and Pfluger 1998). Early fossil evidence for
prokaryotes in lava beds has been dated to a time around 3.5 billion years
ago (Furnes et al. 2004); however, it is unclear exactly when the functional 
genome size reached the present-day minimum value of around $5 \times 10^5$;
the uncertainty in this time value could easily be of the order
of 1 billion years.

Within-group standard deviations in the dates at which respective functional
genome sizes were attained therefore have an order of magnitude spread in
range of values, from 100 million to 1000 million years, with most standard
deviation values being of the order of a few hundred million years. In a
homoscedastic model of the type assumed in the present article, we will use
as a rough (weighted) estimate a value of $s_w \sim 500$ million years.


Therefore, with $\sigma_b \sim 1.36$ billion and $\sigma_w \sim 0.5$ billion, 
we have $\lambda \sim 1.14$ . For uncertainties around 100 million years,
$\lambda$ is around 1.00, and it is around 1.54 if uncertainty is around 
1 billion years.  

\section{Life}

The above formulas for the prediction interval depend on the quantity $\lambda$
Here, we examine
the range of $\lambda$ values which lead to conclusions consistent with
the hypothesis that life did not begin prior to the formation of the Earth.

Figure 2 shows all of the results. The black line shows the ordinary
least-square fit to the data. It is the $x$-intercept of this line, i.e., 
about 9.5 billion years, which led SG to conclude that
life must have begun prior to the formation of the Earth (i.e., about 
4.5 billion years ago). The region between the black, dashed lines is
the 95\% prediction interval for the ordinary least squares fit. According
to this prediction interval (without taking measurement errors into account) 
life may have originated as early as 7 billion years ago. 

\centerline{\includegraphics[height=4in,width=4in]{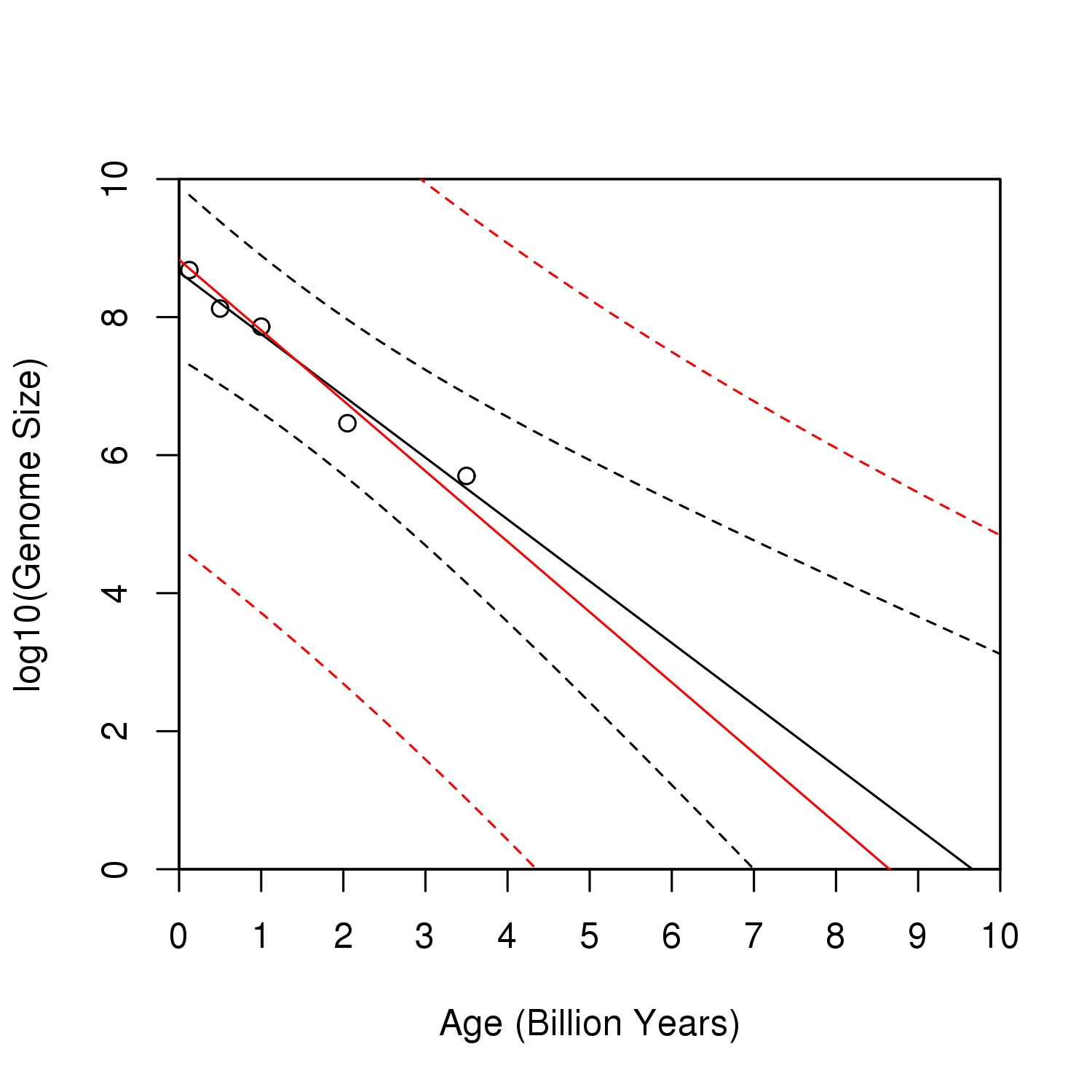} }

The results based on the above measurement error model are shown in red.
The value of $\lambda$ for this fit is 1.14 - the value estimated in
the previous seciton. Note that the resulting prediction interval 
includes 4.5 billion. In other words, if $\lambda$ is about 1.14, then the 
results of the analysis do not reject the hypothesis that life began after 
the formation of the Earth. Even a $\lambda$ as small as 1.1 leads to
results (not shown here) consistent with 4.5 billion years as the age of life. 

Although, a proper interpretation of prediction intervals correctly draws 
all focus away from the ``center" of the interval, it is possible to arrange 
for the corrected fit itself to have an $x$-intercept of 4.5. The result is not
shown here, but the corresponding value of $\lambda$ is about 2.7.
Values of $\lambda$ in the 1.1 to 2.7 range are not uncommon;
Frost and Thompson (2000) even consider $\lambda$ values as large as 5.
More importantly, that range includes the $\lambda$
values estimated in the previous section. 

\section{Conclusion and Discussion}

Recently, Sharov and Gordon (2013) presented an argument to support 
the claim that life must have originated prior to 4.5 billion years ago,
i.e., prior to the formation of the Earth. Here we have shown that an
analysis of the same data allows for life to have been formed
more recently than 4.5 billion years ago, when one takes into account
measurement errors and prediction intervals. Although, we do
not estimate the measurement errors, we demonstrate that the range
of such errors is within the acceptable range. In short, the data
analyzed by SG provide no evidence to reject the hypothesis that 
life formed after the formation of the Earth.

The original conclusion of Sharov and Gordon (2013) is a consequence
of an incomplete analysis. Although the analysis presented here is more 
complete, many improvements are possible. For instance, nonlinear fits
can estimated, and more refined measurement error models can be
developed. The inference component of our analysis - i.e., the 1.96
appearing in the prediction interval - can also be improved upon; the prediction
intervals computed here are only an approximation. This
was sufficient because the main goal of the paper has been to 
introduce measurement error models and to illustrate the importance of 
producing interval estimates (as opposed to point estimates) of the $x$-intercept.  
In short, many aspects of the above formulation are simplistic, approximate,
or even controversial. As such, they offer avenues of further research. 
One aspect, however, that is incontrovertible is that measurement errors
lead to biased (i.e., over- ) estimates for the age of life, and that the bias 
can be corrected/removed. Another unquestionable issue is that all
estimates should be accompanied by some measure of uncertainty (e.g., 
prediction intervals), because without such measures, the conclusions will
not be statistically sound. 

These two recommendations - that measurement error models and interval estimates 
should be employed - are the main lessons of the 
current paper. The details of the measurement model and/or how the
interval estimates are generated are of secondary importance because
they affect our conclusions only in degree, not in kind. However, this analysis
can be improved in a number of ways. The assumption of homoscedasticity can
be relaxed, the $\sigma_w$ and the $\sigma_b$ can be estimated without
the large-$n$ assumption, nonlinear fits can be examined, and one can
even compute a prediction for the $x$-intercept itself. Lastly,
for a more reliable conclusion data sets considerably
larger than used by SG can be employed as genome size data is readily available
for many major transitions along the tree of life (see, e.g.,
{\tt http://commons.wikimedia.org/wiki/File:Tree\_of\_life\_with\_genome\_size.svg}).
\\

\centerline{ {\bf \Large References}}
\begin{description}

\item Bland, J. M., and D. G., Altman, 1994: Statistic Notes: Regression 
towards the mean. {\it British Medical Journal}, {\bf 308}, 1499.

\item Buonaccorsi, J. P., 1995: Prediction in the Presence of Measurement Error: 
General Discussion and an Example Predicting Defoliation. {\it Biometrics},
{\bf 51(4)}, 1562-1569.

\item Buonaccorsi, J. P., 2010: {\it Measurement Error: Models, Methods, and 
Applications}. Chapman \& Hall/CRC Interdisciplinary Statistics

\item Dawkins, R. 2005: {\it The Ancestor's Tale}. Mariner Books.

\item Draper, N. R., and H. Smith 1998, {\it Applied Regression Analysis},
Third Edition. John Wiley and Sons, Inc.

\item Frost, C., and S.G. Thompson, 2000: Correcting for regression dilution bias: 
comparison of methods for a single predictor variable. {\it J. R. Statist. Soc. A},
{\bf 163}, Part 2, 173-189. 

\item Fuller, W. A., 1987: {\it Measurement Error Models}. New York: John Wiley.

\item Furnes, H., N. R. Banerjee, K. Muehlenbachs, H. Staudigel, and M. de Wit, 
2004: Early Life Recorded in Archean Pillow Lavas. {\it Science}, {\bf 304}, 578­581.

\item Galton, F., 1886: Regression towards mediocrity in hereditary stature.
{\it Journal of the Anthropological Institute}, {\bf 15}, 246-263.

\item Montgomery, D. C. 2009: {\it Design and Analysis of Experiments}
(7th Edition), John Wiley \& Sons.  

\item Perrin, E., 1904: On Some Dangers of Extrapolation. {\it Biometrika},
{\bf 3}, 99-103.

\item Rose, K., 2006: {\it The Beginning of The Age of Mammals}.
The Johns Hopkins University Press.

\item Ryan, T. P., 1997: {\it Modern Regression Methods}. John Wiley \& Sons. NY

\item Seilacher, A., P. K. Bose, and F. Pfluger, 1998: Triploblastic Animals More 
Than 1 Billion Years Ago: Trace Fossil Evidence from India" {\it Science},
{\bf 282}, 80-83.

\item Sharov, A. A., 2006: Genome increase as a clock for the origin and 
evolution of life. {\it Biology Direct}, {\bf 1}, 17.

\item Sharov, A. A., and R. Gordon 2013: Life before Earth. 
arXiv:1304.3381 [physics.gen-ph]

\end{description}

\end{document}